**Expert Programming Knowledge: a Schema-Based Approach**

Françoise Détienne
Projet de Psychologie Ergonomique pour l'Informatique
INRIA
Domaine de Voluceau,
Rocquencourt, BP 105
78153, Le Chesnay Cedex
France
Email address: detienne@psycho.inria.fr
Phone Number: (1) 39 63 55 22

The topic of this chapter is the role of expert programming knowledge in the understanding activity. In the "schema-based approach", the role of semantic structures is emphasized whereas, in the "control-flow approach", the role of syntactic structures is emphasized. Data which support schema-based models of understanding are presented. Data which are more consistent with the "control-flow approach" allow to discuss the limits of the former kind of models.

The structures of knowledge and their organization are important characteristics of a model of the expert programmer. Having such a model allows to design systems which support programming activities. Such a model is also necessary to understand how expertise is built in a domain. The topic developed in this chapter is the role of expert knowledge in programming activities. For two reasons, only the activity of understanding is developed. First, as shown in Pennington's chapter, the same kind of knowledge is used by the processes of program composition and program comprehension. Second, the understanding activity is particularly interesting to analyze because it is involved in several maintenance tasks. Constructing a representation of a program is effectively necessary for debugging or modifying a program. Now these tasks correspond to the main part of programmers' activity.

Many studies on programming knowledge have been conducted in the theoretical framework of Schema Theory. This approach is developed in this chapter and is evaluated on the basis of empirical data.

In section One, studies on program understanding are briefly reviewed from a historical perspective. In the "schema-based approach", the role of semantic structures is emphasized whereas, in the "control-flow approach", the role of syntactic structures is emphasized. In section Two, the knowledge organization which programmers are supposed to possess is developed. Then, empirical support to hypotheses on knowledge structures is presented. In section Three, understanding mechanisms are analyzed through empirical results.

## A HISTORICAL PERSPECTIVE ON APPROACHES OF PROGRAM UNDERSTANDING

Different approaches can be distinguished according to the kind of knowledge and the kind of mechanisms to use this knowledge which are assumed to be involved in experts program understanding. Authors disagree on the kind of knowledge which is important to understand a program.

In the schema-based approach (as called in this chapter) experts are assumed to use mainly knowledge structures which represent semantic information in the programming domain. These structures group together information on parts of programs which are involved in performing a same function. In the control-flow approach experts are assumed to use mainly knowledge structures which represent elements of the control structure. These structures group together parts of programs of the same syntactic structure. The former approach emphasizes the role of semantic knowledge, i.e., knowledge on what the program does whereas the latter approach emphasizes the role of structural knowledge, i.e., knowledge on how the program functions.

From a theoretical viewpoint, authors of the schema-based approach have used concepts developed in the Schema Theory. Knowledge structures in programming are formalized in terms of "programming schemas" or "programming plans". The Schema Theory is not always referred to in the "control-flow approach". However the concept of schema may also be used in modeling the syntactic knowledge structures experts are assumed to possess.

The Schema Theory is a theory on knowledge organization in memory and on processes involved in using knowledge. A Schema is a data structure which represents generic concepts stored in memory. This theory has been developed in Artificial Intelligence (Abelson, 1981; Minsky, 1975; Schank and Abelson, 1977) and in psychological studies on text understanding (Bower et al., 1979; Rumelhart, 1981).

The schema-based approach in programming studies has begun with Rich's work (1981) and has been mainly developed by the group headed by Elliot Soloway at Yale University. It has allowed to account for problem solving and understanding activities involved in various programming tasks such as program design, debugging, enhancement. The studies have started in the eighties and some of the most important papers on this topic certainly are (Soloway et al., 1982a; Soloway et al., 1982b; Soloway and Ehrlich, 1984). For several years, others researchers from different institutions (Brooks, 1983; Détienne, 1989; Rist, 1986) have conducted studies on programming in the same theoretical framework. In the "control-flow approach", studies have also been developed both by computer scientists and psychologists (Atwood & Ramsey, 1978; Curtis et al. 1984; Linger et al. 1979; Pennington, 1987; Shneiderman, 1980).

Authors not only disagree on the kind of knowledge which is important to understand a program. Authors also disagree on the way knowledge is used in program understanding. On one side, authors from the schema-based approach assume that understanding a program consists in evoking a programming schema (or several schemas) stored in memory, instantiating that schema with values extracted from the text, inferring others values on the basis of the evoked schema. The mechanisms of schemas activation can be either data-driven or conceptually-driven: in the first case the activation spreads from substructures to superstructures,and, in the second case, the activation spreads from the superstructures to the substructures. So, semantic structures can be evoked directly by information extracted from the code or by other activated schemas.

On the other side, authors from the control-flow approach assume that understanding consists in identifying control structures then combining these structures into larger structures until identifying groups of structure corresponding to functions. A model of program understanding, the syntactic/semantic model, has been developed by Shneiderman and Mayer (in Shneiderman, 1980) in this state of mind. They assume that understanding activity involves two separate processings: a syntactic processing which occurs first and a semantic processing occurring afterwards.

In the following of this chapter, the "schema-based approach" is mainly developed. Data which support schema-based models of understanding are presented. However, data which are more consistent with the "control flow approach" are also presented; they allow to discuss the limits of the former kind of models.

## KNOWLEDGE ORGANIZATION IN MEMORY

### 1. Theoretical framework

In the schema-based approach of understanding hypotheses are made on the typology of schemas possessed by experts, the relationship between schemas, and the structure of representations constructed from programs. These points are developed in this section. Remarks are also made on the knowledge organization assumed in the control-flow approach.

*Typology of schemas possessed by experts*

Soloway et al. (1982a) have encoded experts' schema knowledge as frames. A schema is represented as a knowledge packet with a rich internal structure. A schema (or "plan" in their terminology[1]) is composed of variables ("slot types") which can be instantiated with values ("slot fillers").

Program understanding involves the evocation of schemas of different domains and their articulation. Brooks (1983) assumes that program understanding involves a mapping between, at least, two domains, the programming domain and the problem domain. Experts would possess schemas representing information on problems which are dependent on the problem domain (or task domain). Détienne (1986b) describes the general structure of those schemas as including slots on the data structure and on the possible functions in a particular problem domain. For the particular domain of stock management, the following values could be assigned to the slots:

>  Task domain: stock management
>  Data structure: record (name of file, descriptor of file..)
>  Functions: allocation (creation or insertion), destruction, search

Among the schemas relative to the programming domain, Soloway and al. distinguish variable plans and control flow plans. Variable plans allow to generate a result which is stored in a variable. For example, a Counter_Variable plan can be formalized as:

>  Description: counts occurrences of an action
>  Initialization: Counter:=0
>  How used? (or update): Counter:=Counter+1
>  Type: integer
>  Context: iteration

The role of control flow plans is not to generate results, but to regulate the use and production of data by other schemas. For example, a Running_Total_Loop plan which allows to compute the sum of numbers involves initializing a total variable to zero, telling the user to enter a number, reading that number and looping until a stopping value is entered. It is described by Soloway as the following frame:

>  Description: build up a running total in a loop, optionally counting the numbers of iterations
>  Variables: Counter, Running_total, and New_Value
>  Set up: initialize variables
>  Action in body: Read, Count, Total

In this schema, the Running_total variable refers to a variable which builds up a value one step at a time, like, for example, a variable in which would be accumulated the sum of read numbers. The New_Value variable allows to hold values produced by a generator. This can be, for example, a Read_Variable plan which receives and holds a newly read variable or a Counter_Variable plan which counts occurrences of an action. Experts would also possess more complex schemas representing algorithms like search or sort algorithms or abstract types like tree structures or record structures.

Pennington (1987) describes different kinds of relations which formally compose a program: control flow relations which reflect the execution sequence of a program, data flow relations which reflect the series of transformation of data objects in a program, functional relations which concern the goal achieved in a program. So as to contrast the schema-based approach and the control-flow approach, this author remarks that Plan knowledge analysis is closely related to an analysis of program text in terms of data flow relations and functiona relations. Programming schemas represent information relative to functions performed in a program such as search an item and they are based on data flow relations. On the contrary, in the control-flow approach, knowledge analysis is more closely related to elements of the control structures which are sequences, iterations and conditional structures. Those basic structures of programs are called "Primes structures" by Linger, Mills and Witt (1979).

---

[1] I will indifferently use the terms "plan" and "schema" in the following of this chapter.

Soloway et al. assume that experts also possess rules of discourse which are programming conventions on the way to compose and to use programming plans. For example, a rule expresses that the name of a variable should reflect its function. The application of these rules produces prototypical values associated to plans.

*Relationship between schemas*

Soloway et al. (1982b) describe two kinds of relationship between plans: a relation of specialization ("a kind of") and a relation of implementation ("use").

The relation of specialization links together plans which are more or less abstract. A specialized plan forms a subcategory of the plan just above it in the hierarchy. For example a New_Value_Variable Plan which holds values produced by a generator can be specialized in two distinct plans: a Read_Variable plan and a Counter_Variable plan. In the same way, the Running_Total_Loop plan can be specialized as three distinct plans: Total_Controlled_Running_Total Loop plan in which the Running_Total is tested, a Counter_Controlled Running_Total Loop plan in which the Counter is tested and a New_Value_Controlled_Runnning_Total_Loop plan in which the New_Value is tested.

The relation of implementation links together plans which are language independent on one side, with plans which are language dependent on the other side. Implementation plans specify language dependent techniques for realizing Tactical and Strategic Plans. For example, the For_Loop plan is a technique for implementing the Counter_Controlled Running_Total_Loop plan in Pascal.

The relation of implementation refers also to the way some plans are composed of elementary plans. For example a Running_Total_Loop Plan is composed of different Variable plans which are associated to it by the use link.

*Program representation*

In the schema-based approach, program may be represented as composed of goals/subgoals with programming schemas associated to goals so as to achieve them. Fig. 1a illustrates the notions of goals and subgoals for a program which computes the average of read numbers. Goals are decomposed in subgoals. For example, the goal "report-average" is decomposed in three subgoals "enter-data", "compute-average", "output-average". At each terminal subgoal in the tree is associated a schema used to realize it in a Pascal program.

INSERT FIGURE 1 ABOUT HERE

Fig. 1b illustrates the implementation of schemas in a program. It shows, for example, that the variable whose name is "Count" implements a Counter_Variable plan. Its initialization corresponds to the line "Count:=0" and its update corresponds to the line "Count:=Count + 1".

In the control-flow approach, a program representation would be structured, at a detailed level, in terms of syntactic structures which could be combined, at a more abstract level, in terms of semantic structures.

## 2. Empirical support

Empirical data support the hypothesis that experts possess schemas of programming which represent semantic information. Different kinds of support to the schema-based approach are provided by studies, using categorization tasks and using fill-in-the-blank tasks for plan-like and unplan-like programs.

*Categorization of programs*

The experts' knowledge organization gives them a capacity of processing which is superior to novices. Experts recall programs better than novices whenever the order of presentation is correct (Shneiderman, 1976) but this superiority disappears when programs are presented in disorder so that meaningful structures do not appear.

The categories formed by experts the knowledge structures they use. The categories formed by novices should be different from those formed by experts inasmuch as novices do not possess the same knowledge structures. Using a recall procedure, Adelson (1981) shows that categorization of programs or parts of programs is different according to the expertise of subjects. Novices' categories depend on surface structures of the program like syntactic structure; on the contrary, experts categories cluster, on one hand, the elements of programs performing a same function and, on the other hand, elements of programs displaying a procedural similarity .

These last results suggest that experts possess knowledge structures grouping together information relative to a same function. This supports the hypothesis of programming plans. However, Adelson's results also suggest that experts possess knowledge structures which reflect the control structure of the program. This last finding is more consistent with the control-flow approach.

*Understanding plan-like and unplan-like programs*

Soloway and Ehrlich's results (1984) support the hypothesis that experts possess and use programming plans and rules of discourse in comprehending programs. If this hypothesis is correct, then giving experts programs which disrupt plan structure should make them much more difficult to understand than programs which do not. On the contrary, novices which do not possess programming plans and conventions yet should not be sensitive to the fact that the programs do or do not conform to programming plans.

To evaluate this hypothesis, Soloway and Ehrlich have constructed two versions of programs: plan-like versions and unplan-like versions. In unplan-like versions, the way the plans are composed is not prototypical (usual), or a plan is implemented with a value which is not prototypical. A way of constructing unplan-like programs is to construct programs that violate some rules of discourse. Fig. 2 presents an example of plan-like and unplan-like versions for a program which computes an average.

INSERT FIGURE 2 ABOUT HERE

The subjects performed a fill-in-the-blank task: one line of the code was erased and the subjects had to fill in the blank line with a line of code that in their opinion best completed the program. They were not told what the program was supposed to do. The results are that the experts performed better than the novices, that the subjects answered correctly the plan-like versions more often than the unplan-like versions, and that there is an interaction between program version and expertise. As expected, the decrease of performance between unplan-like and plan-like versions is more important for experts than for novices. Those results support the hypothesis on experts' plan knowledge structures.

Furthermore the schema-based model predicts which kind of errors experts can do in unplan-like condition. If their understanding is based on programming plans and rules of discourse then they will try to infer the missing line on this basis. Thus they should tend to give plan-like answers even for unplan-like versions. This was observed in the Soloway et al. experiment.

# UNDERSTANDING MECHANISMS

## 1. Theoretical framework

According to a schema-based approach of understanding, schemas representing semantic knowledge are evoked while reading a program. They may be evoked either in a bottom-up way or in a top-down way. The direction of activation is bottom-up when the extraction of cues from the code allows the activation of schemas or when an evoked schema causes the activation of schemas it is part of. The direction of activation is top-down when evoked schemas cause the activation of less abstract schemas; activation spreads from superstructures toward substructures. In Brooks'model (1983), the activation process is assumed to be mostly conceptually-driven.

Concerning the evocation of schemas, Brooks attributes an important role to "beacons" that is to features or details visible in the program or the documentation as typical indicators of the use of a particular operation or structure. They allow the activation or recognition of particular schemas.

Détienne (1989) stresses that as most of schema-based models focus on the construction of goal/plan representation based on activation and instantiation processes, it seems important to analyze the processes which allow to evaluate this representation. Détienne distinguishes two processes: the evaluation of the internal coherence between plans and the evaluation of the external coherence between plans. The former consists in checking whether or not the values instantiated in a plan satisfy the constraints on the instantiation of the plan's slots. The latter consists in checking whether or not there are interactions between plans and between goals and if they create any constraints on the implementation of plans.

In the schema-based approach, the programmer is assumed to construct a representation of the functions performed in the program, making more or less explicit the data flow relations in the program.

In the control-flow approach, syntactic constructs are assumed to be evoked first. That means that the programmer construct a representation of the control flow relations in the program before constructing a representation of the performed functions. Syntactic constructs are assumed to be combined in more and more abstract constructs until reaching a level of functional representation.

## 2. Empirical support

Data on inferences a programmer makes in understanding tasks and recall tasks reflect the kind of knowledge which is evoked by the subject in the understanding activity and give information on the way knowledge is used. Data will be presented from different studies. "Chunks" which the programmer perceive in the code of a program read group together information belonging to the same mental constructs. Empirical data give us information on the kinds of chunks constructed by experts. The kind of evoked knowledge and understanding mechanisms may vary according to the task performed. Data are given on the activity developed in different tasks of maintenance.

Observations support the schema-based approach whereas others observations support the control-based approach. This will be discussed in the last section.

*Inferences collected in understanding tasks*

According to the schema-based approach, schemas representing semantic knowledge are evoked in program understanding. These schemas allows to draw inferences. Inferences collected in understanding tasks should reflect the kind of knowledge which has been evoked in memory. To study this process of schema activation, Détienne (1986a; 1988) designed an experimental setting in which experienced programmers had to verbalize while reading programs (written in Pascal) presented one instruction at a time. At each instruction newly presented, the subjects had to tell the information provided by the new instruction and the hypotheses they could elaborate concerning the other instructions and the functions performed in the program.

The results support the hypothesis that experts possess programming plans like those formalized by Soloway. The verbal and behavioral protocols have been coded in the form of production rules: "IF A, THEN B". These rules represent various mechanisms for using knowledge which allow subjects to draw inferences. Evoked knowledge structures have been formalized as frames "SCHEMAn" composed of variables "VARn". 269 rules have been identified which have been classified as examples of 47 general rules. In its general form, a rule describes the types of variables (slot types) composing a schema. In its instantiated form, it describes possible values (slot fillers) which allow to instantiate a variable. In this section, several examples of these rules are presented in their instantiated form, with the slot types in brackets.

Some identified rules describe activation processes which are data-driven (the direction of activation is bottom-up); they are expressed as "IF VAR11, THEN SCHEMA1" (VAR11 is a variable composing SCHEMA1) or as "IF SCHEMA11, THEN SCHEMA1" (SCHEMA11 is linked to a variable composing SCHEMA1). As schema instantiation begins as soon as a schema is evoked, some rules express also instantiation processes; this is expressed as "IF VAR11, THEN VAR12 (VAR11 and VAR12 are two variables which are part of the same schema).

Détienne observed that a Counter_Variable plan, for example, can be evoked by the extraction of cues such as the variable's name or its form of initialization. This is illustrated by the following rules:

```
IF VAR11 (variable's name): I
IF VAR12 (variable's type): integer
THEN SCHEMA1 (schema of variable): Counter_Variable plan
THEN VAR13 (context): iteration

IF VAR11 (variable's initialization form): I:=1
THEN SCHEMA1 (schema of variable): Counter_Variable plan
THEN VAR12 (variable's update): I:=I+1
```

These rules make conspicuous that the experts infer other values associated to a Counter_Variable plan when it is activated. According to the first rule, the reading of the declaration of a variable whose name is "I" and type is "integer" allows to evoke a Counter-variable plan. This activation allows the programmer to infer the value associated to another slot of this schema which is the context in which the variable "I" is used. So, the subject expects to see some kind of iteration in the code of the program. According to the second rule, the activation of a Counter-variable plan allows the subject to expect a particular form of update which is an incrementation .

Détienne observed that activation of elementary schemas such as variable schemas can allow the evocation of schemas representing algorithms which are partly composed of those elementary schemas. This is illustrated in the following rule.

```
IF SCHEMA11 (schema of variable): Counter_Variable plan: initialization I:=Z, name I
IF SCHEMA 12 (schema of loop): while A<>B
THEN SCHEMA1 (algorithmic schema): Linear_Search plan
THEN SCHEMA 11 (schema of variable): Counter_Variable plan: update: I:=I+1
```

In this example, the form of initialization and the name of a Counter variable and the form of a loop (while) allow the evocation of an algorithmic schema which is a schema for linear search. The elementary schemas which allow this activation are a counter variable for which subjects infer the update and a schema of loop.

Other rules describe activation processes which are conceptually-driven (the direction of activation is top-down); this is expressed as "IF SCHEMA1 and..., THEN SCHEMA2" (SCHEMA 2 is a subcategory compared to SCHEMA1) or as "IF SCHEMA1, THEN SCHEMA11, THEN SCHEMA13, ..., THEN SCHEMA1n" (the SCHEMA1ns are linked to variables which compose SCHEMA1). Détienne (1988) gives an example of this process which allows the expectation of a complex combination of several schemas implemented in the code.

Norcio results (Norcio, 1982) also show that semantic constructs are evoked and allow to draw inferences while reading a program. Expectations can be drawn from comments in program understanding. He asked subjects to fill in a blank line either at the beginning of a chunk or in the middle of a chunk (or groupings of code corresponding to cognitive units). In an experimental setting, comments were inserted in the program right before cognitive units. Results show a positive effect of comments compared to no comments in a fill-in-the-blank task when the line to fill in is at the beginning of a chunk. In that case, the comment provides the subjects with a cue for schema evocation.

Widowski and Eyferth (1986) have collected data on reading strategies used by programmers for programs varying along a dimension of stereotypness. They remark that the strategies used by experts are different for usual and unusual programs. When reading usual programs, the activity seems to involve a conceptually-driven processing. When reading unusual programs, they describe a more bottom-up oriented processing.

It is noteworthy that, in some cases, the evocation of schemas may create negative effects on the performance inasmuch as the presence of inferred values is not checked in the code. Détienne (1984) observed this kind of negative effect in an experiment in which experts had to realize a debugging task on a program written by somebody else. She remarks that the experts have difficulties to detect some errors when they strongly expect a particular value in the code (this value being different and incorrect in the code) and they do not verify whether or not the value is actually present. This kind of negative effect is more likely to happen when the program has been written by the reader himself/herself since he/she has strong expectations on what should be in his/her program.

In summary, data support the hypothesis that programming plans are evoked in program understanding. The evocation of schemas allows to draw inferences. However, it may happen that there is no cues in the code to evoke schemas. Mental execution has been shown to be used to infer the goal of a part of code when the programmer has no expectation about it. By acquiring knowledge about the intermediate values the variables take during execution, he/she can infer the goal of the process, and so, the goal of the part of code they have executed. Experts debugging programs have been observed to mentally simulate while not having enough textual cues about the goal of a part of code: no name, no familiar structure, no documentation (Détienne, 1984). This result suggests that schema evocation may be based either on static information like "beacons" or "dynamic information" which change with the program execution. This last kind of information is closed to data flow relations in a program.

Détienne and Soloway (1988) remark that concrete simulation is also used to infer information about the interaction between plans. In an experiment in which experts had to perform a fill-in-the-blank task on programs and had to verbalize while performing that task, they show that whatever the planliness of programs is, the experts use concrete simulation when they want to check for unforeseen interactions. This is typically used to understand programs which compute an average: what is specific to those programs is that a loop is used with a counter. This suggests that the experts know by experience that the use of a count plan in a program causes unforeseen interactions and that the best way to check for these interactions is to execute the part of the program with the count.

*Inferences collected in recall tasks*

Inferences collected in recall tasks subsequent to understanding tasks reflect the kind of knowledge which has been evoked in memory. Results of studies support the hypothesis of semantic structures whereas other results support the hypothesis of syntactic structures.

Evidences support the schema-based approach. In two experiments conducted by Détienne (1986b), experts were asked to recall the program read after conducting a debugging task. During the reading phase, the subjects had to discover an unknown program and to evaluate its correctness. The programs were written in Pascal.

Distortions of the form of the program were observed in the recall protocols. For example, distortions concerned the name of a variable used as a counter. The subjects recalled I instead of J. This suggests that the variables are memorized in a category, a Counter_Variable plan, and that the lexical form is not kept in memory. In the recall process, the subjects use another possible value of the slot "name of variable".

Other observations suggest the existence of prototypical values in the slots of a schema. Each slot is associated with a set of possible values, as seen before. Those values have not the same status, some values being more representative of a slot for a schema than the others. When there is a prototypical value in a category, this value comes first in mind when the category is activated.

In a schema for a flag, the slot "context" is an iteration that can take the values "repeat" or "while". In the program used, this variable,V, appears in a "repeat" iteration. During the reading phase, the value of the iteration expected by most of the subjects was "while not V do". In the recall protocols, a distortion was observed: a subject has recalled the instruction as "while" instead of "repeat". This subject has reported the prototypical value instead of the adequate value. These observations give support to the hypothesis of prototypical values being associated to the slots of the schemas.

Distortions of the content of the program were also observed in the recall protocols. When a schema is activated, information associated to this schema is inferred. So information typical of a schema may be recalled while not included in the program text. Détienne reports a thematic insertion which was observed. For the particular domain of stock management, the following values could be assigned to the slots of the problem schema :

    Task domain: stock management
    Data structure: record(name of file, descriptor of file)...
    Functions: allocation (creation or insertion), destruction, search

The function of creation was not isolated in a subprogram and the subjects did not read any information concerning this function in the program. Nevertheless, a subject has reported this function as if it was a subprogram. This suggests that schematic knowledge dependent on the task domain has participated to the elaboration of the representation and to the process of recovering of stored information at recall.

So, data suggest that the knowledge structures evoked in understanding are organized according to the semantic of information. However, other data suggest (1) that experts also use knowledge structures which are organized in function of syntactic information (procedural like) and (2) that a representation based on this kind of knowledge may be constructed previously to a representation constructed on the basis of semantic knowledge (functional like). This is compatible with the control-flow approach.

In Pennington's study (1987), subjects were asked to read a short program before answering questions. Results show that questions about control flow relations are answered faster and more correctly than questions about data flow and functional relations. Furthermore, in a recognition memory test, subjects recognized a statement faster when it was preceded by another segment of the same control structure than when it was preceded by a segment of the same functional structure. This suggests that knowledge structures representing control flow have an important role in program understanding. Results also suggest that the understanding of the program control structures may precede the understanding of program functions.

However, results also suggest that understanding may be schema-based or control-based according to the understanding situation. Pennington remarks there is an effect of the language. Two languages were used in this experiment: Fortran and Cobol. Cobol programmers were better at responding to comprehension questions about data flow than were Fortran programmers and control flow relations were less easily inferred by Cobol programmers. Pennington also remarks there was an effect of the comprehension task, i.e., the goal the subjects have in understanding the program. After a modification task, Pennington notices there was a mark shift toward increased comprehension of program function and data flow at the apparent expense of control flow information.

*Chunks constructed in program understanding*

A chunk is the result of the identification of units belonging to a same mental construct. So a chunking task allows to study the cognitive units on which programmers base their comprehension. Several studies highlight that the chunks constructed by experts and novices in understanding are different and that experts' chunks reflect semantic structures. This is both compatible with the schema-based approach.

As programmers analyze programs on the basis of units which correspond to cognitive constructs, it is likely that highlighting those units in programs would facilitate the understanding process. Norcio (1982) has shown this type of effect by indenting programs on the basis of functional chunks defined by programmers and asking subjects to fill in a blank line either at the beginning of chunks or in the middle. The results indicate that subjects with indented programs supply significantly more correct statements compared to the non-indented group.

Black et al. (1986) and Curtis et al. (1984) remark that elements of code which are parts of the same plan are often dispersed in a program. So are the initialization and the updating of a Counter-variable plan. This characteristic of interleaving (see Green's chapter) would be language dependent. This dispersion would make these parts difficult to integrate into a functional whole. This characteristic would be part of what makes program understanding hard. Letovsky and Soloway (1986) illustrate how difficult a program is to understand when using what they call a "delocalized plan", i.e. a plan whose parts are dispersed in the program.

Rist (1986) provided experts and novices with programs to describe in terms of groups of lines of code which "did the same thing". In describing a program, novices used a mixture of syntactic and plan-based chunks. Experts used almost only plan-based groupings. Rist notices that when programs are complex, plan use decreases. In that case, construction of program representation can not be made from plan structure only, thus subjects use control-based program understanding. This last finding suggests that understanding may be schema-based or control-based according to the understanding situation.

*Understanding mechanisms in different tasks*

The understanding activity is involved in different tasks of maintenance. Two tasks have been studied: the enhancement (or modification) task and the debugging task. In both of these tasks, studies show that experts may evoke programming plans so as to construct a representation of the program. However, studies also stress the importance of the simulation mechanisms. This highlights the role of information on data flow relations and control flow relations in these tasks. This is compatible both with the schema-based approach and the control-flow approach.

In the task of modification, mental simulation has been observed to be used by experts. As simulation allows to infer information on connections and interactions between functional components, it is likely to be used in tasks like enhancement in which processing this kind of information is particularly useful. Results from Littman et al. experiment (1986) show that some experienced programmers use this symbolic execution so as to acquire causal knowledge which permits them to reason about how the program's functional components interact during reading. Symbolic simulation is used to understand data flow and control flow.

This kind of strategy is the most effective inasmuch as it prevents subjects from introducing errors by modifying a part without taking into account the relationship between that part and other parts in the program. Simulation is a way to evaluate the external coherence between plans, i.e. to check whether or not there are interactions between plans and between goals (Détienne, 1989).

As reported before, Pennington (1987) notices there was a mark shift toward increased comprehension of program function and data flow at the apparent expense of control flow information after a modification task.

Concrete simulation, i.e., executing the program with values, is particularly useful in tasks like debugging in which it is important to judge whether the values produced at the execution are the expected values. In the debugging task, experts have been observed to mentally simulate the program so as to have information on the values taken by variables during the program execution (Détienne, 1984; Vessey, 1985).

It is noteworthy that mental simulation has also been observed to be involved in design tasks as explained in Visser and Hoc chapter. It is not surprising however, as design task involves some understanding activities. It is used so as to predict potential interactions between elements of the design and to check parts of the programs.

## DISCUSSION

Empirical data support the schema-based approach of program understanding. However data also support the control-based approach. Studies show that according to the understanding situation, knowledge used may be related to different kind of information: data flow relations, functional relations or control-flow relations. Programming plans formalize information on data flow and functions whereas syntactic constructs reflect more the structure of the program as described with control flow relations.

However, the two approaches of understanding are not contradictory. First, as said before, syntactic constructs may also be formalized as schemas. Second, a model of expert' knowledge may integrated those different kinds of knowledge. Thus, it should characterize the understanding situation so as to account for what kind of knowledge is used, when it is used and how it is used.

The understanding situation may be described by the characteristics of the language, the task, the environment and the subject. Concerning the language, the presence of cues in the code which allow the activation of schemas representing semantic knowledge may be dependent on the notational structure. Green (see Green's chapter) assumes that languages vary along a dimension of "role-expressiveness". With a "role expressive" language, the programmer can easily perceive the purpose, or role, of each program statements . So, schemas evocation may be based on static cues whereas, with non "role expressive" languages, it may be based on the extraction of dynamic information like control-flow information.

Studies of programming have been developed on the basis of experiments conducted with a relatively narrow sample of programming languages: mostly Pascal, rarely Lisp, more rarely Basic, more recently Prolog. It seems important now to conduct studies with other languages (for example, object-oriented languages) so as to take into account the effect of languages' characteristics in the understanding activity.

A model of the expert should be extended to take into account task variations. As Gilmore and Green (1984) remark, the information needed to be extracted from the code is different according to the task in which the programmer is involved. Furthermore, a particular language may make explicit in the code certain information or a particular environment may emphasize certain information important to achieve a particular task. So knowledge used by the expert may depend on the goal of his/her activity and on the availability of information in a particular situation.

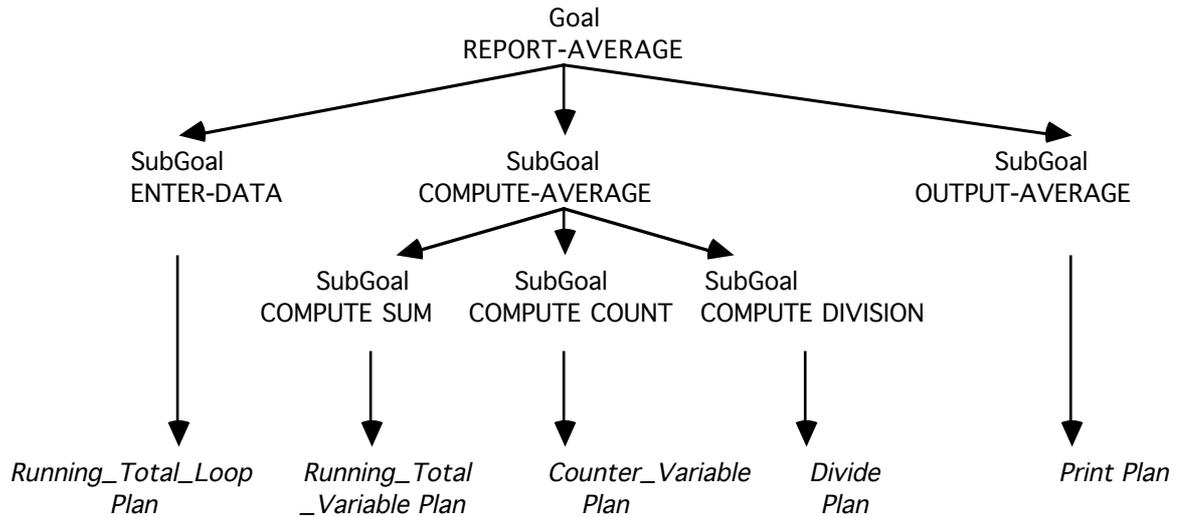

Fig. 1a:
Hierarchical representation of goals and schemas

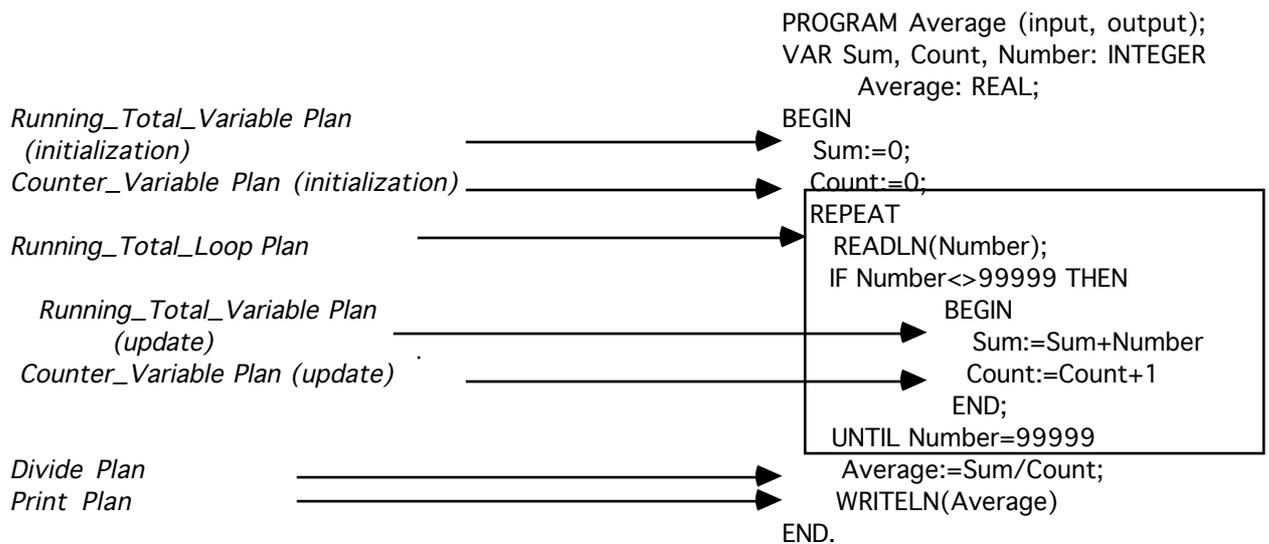

Fig. 1b:
Representation of the combination of schemas implemented in the program code

Fig. 1:
REPRESENTATIONS OF A PROGRAM COMPUTING AN AVERAGE

**PLAN-LIKE VERSION**
PROGRAM Grey (input, output)
Var Sum, Count, Num: INTEGER
   Average REAL
BEGIN
   Sum:=0;
   Count:=0;  *  *line to fill in*
   REPEAT
      READLN(Num);
      IF Num<>99999 THEN
                      BEGIN
                           Sum:=Sum+Num;
                           Count:=Count+1;
                      END;
   UNTIL Num=99999;
   Average:=Sum/Count;
   WRITELN(Average);
END.

**UNPLAN-LIKE VERSION**
PROGRAM Orange(input, output)
VAR Sum, Count, Num: INTEGER;
   Average: REAL;
BEGIN
   Sum:=-99999;
   Count:=-1;  *  *line to  fill in*
   REPEAT
      READLN(Num);
      Sum:=Sum+Num;
      Count:=Count+1;
   UNTIL Num=99999;
   Average:=Sum/Count;
   WRITELN(Average)
END.

**DESCRIPTION** (extract from Soloway and Ehrlich, 1984)
This program calculates the average of some numbers that are read in; the stopping condition is the reading of the sentinel value, 99999.
The plan-like version accomplishes the task in a typical fashion: variables are initialized to 0, a read-a-value/process-a-value loop is used to accumulate the running total, and the average is calculated after the sentinel has been read.
The unplan-like version was generated from the plan-like version by violating a rule of discourse: *don't do double duty in a non-obvious way*. That is, in the unplan-like version, unlike in the plan-like version, the initialization actions of the COUNTER VARIABLE (Count) and RUNNING TOTAL VARIABLE PLANs (Sum) serve two purposes:
-Sum and Count are given initial values
-the values are chosen to compensate for the fact that the loop is poorly constructed and will result in an off-by-one bug: the final sentinel value (99999) will be incorrectly added into the RUNNING TOTAL VARIABLE, Sum, and the COUNTER VARIABLE, Count, will also be incorrectly updated.

FIGURE 2.
Program AVERAGE